# Black Holes:
# Attractors for Intelligence?

Clément Vidal
Center Leo Apostel
Evolution, Complexity and Cognition research group
Vrije Universiteit Brussel (Free University of Brussels)
Krijgskundestraat 33, 1160 Brussels, Belgium
Phone +32-2-640 67 37 | Fax +32-2-6440744
http://clement.vidal.philosophons.com
clement.vidal@philosophons.com

**Abstract:**
The Search for Extra-Terrestrial Intelligence (SETI) has so far been unsuccessful and needs additional methods. We introduce a two-dimensional metric for civilization development, using the Kardashev scale of energy increase and the Barrow scale of inward manipulation. To support Barrow's scale limit, we contend with energetic, societal, scientific, computational, and philosophical arguments that black holes are attractors for intelligence. An application of the two-dimensional metric leads to a simple, consistent and observable hypothesis to test the existence of very advanced civilizations. We suggest that some already observed X-Ray binaries may be unnoticed advanced civilizations, of type KII-B$\Omega$. The appendix provides an argumentative map of the paper's main thesis.

**Keywords:** SETI, black holes, Kardashev scale, Barrow scale, star lifting, XRB

## Contents

# Introduction

My aim in this paper is to introduce a framework to find very advanced extra-terrestrial civilizations. In the Search for Extra-Terrestrial Intelligence (SETI) field, arguments converge showing that presumed extra-terrestrial civilizations are between 1.7 and 8 billion years older than us (see Dick 2003, 67 and references therein). If this is the case, we need *not* to be overcautious in our SETI speculations. Quite on the contrary, we must push them to their extreme limits if we want to glimpse what such civilizations could look like. Naturally, such an ambitious search should be balanced with cautious conclusions. Furthermore, given our total ignorance of such hypothetical civilizations' properties, it seems wise to encourage and maintain a wide variety of search strategies. A commitment to observation and to the scientific method remains our best touchstone.

Frank Drake's (1961) famous equation has inspired much of our understanding of cosmic evolution and helped us to frame agendas for SETI. This equation is a tool to assess "the number of *communicative* civilizations which might exist in our *galaxy*" (my emphasis). However inspiring and helpful it has been, it has also introduced two fundamental biases in SETI.

First, it focuses on *communication*. This is the orthodox way of searching for messages coming from an Extra-Terrestrial Intelligence (ETI). This program has failed so far. One may advance many good reasons for this failure, but the bottom line is that we do not need to assume communication to conduct SETI. The equation introduces a second bias by focusing on our galaxy *only*. By endorsing the Drake equation's agenda uncritically, we study one object, our galaxy, out of the 170 billion ($1.7 \times 10^{11}$) galaxies estimated to shine in the universe (see e.g. Gott III et al. 2005). For more critique on this limiting galactocentrism, see also (Ćirković and Bradbury 2006).The good news is that if we extend Drake's equation to the whole universe, then our detection chances increase. More precisely, looking at old galaxies in the distant past provides the opportunity to test wide-ranging scenarios for civilization development at different periods.

Those two biases have shifted SETI's fundamental question from (1) "Are we alone?" to (2) "Who wants to chat in the galaxy?". Of course, it would be much more enriching and fun to communicate or to have direct contact with ETIs. Accordingly, starting SETI in our own galaxy is also the first logical and practical step to take. Yet, if we really wish to find out whether we are alone or not (1), it requires extending our search strategies.

In this paper, I take a *Dysonian approach* to SETI, emphasizing the search for extra-terrestrial technological manifestations and artifacts (see e.g. Dyson 1960, 1966; Ćirković 2006). This approach is also in line with the more recent framework of the *postbiological universe* introduced by Steven J. Dick, which includes insights from astrobiology, computer science and futures studies (Dick 2003; Ćirković and Bradbury 2006). This framework invites examination of new kinds of objects. For example, Seth Shostak (2010, 1029) recently proposed to widen the search to bok globules (cold molecular clouds), hot stars, neutron stars, and … black holes. Importantly, we do not make any particular hypothesis on the form of ETIs. They can be biological, post-biological, or even based on other life-principles. They can be willing to communicate or not, thriving in our galaxy or in others.

I first introduce a two-dimensional metric for civilization development using, on the one hand, Kardashev's (1964) scale of energy consumption increase; and on the other hand, Barrow's (1998) scale of smaller and smaller scale manipulation abilities (section 1). Exploring the limits of the Barrow scale, I show with a surprisingly wide variety of arguments that black holes are attractors for intelligence (section 2). As an application of the two-dimensional metric, I introduce *black hole star lifting*, offering a fresh SETI perspective on binary systems (section 3). If the reasoning holds, corresponding civilizations (KII-B$\Omega$) to those binary systems may already have been observed without due recognition. Since it is too early to draw any definitive conclusion, I close the paper with some crucial open questions.

# 1 Two scales for civilization development

We can distinguish two very general scales for civilizations development (Table 1). Kardashev's scale measures the energy consumption of a civilization. It has been refined since its original publication, but its original version will suffice for our purpose. Barrow's scale measures a civilization's ability to manipulate small-scale entities. It has been largely ignored up to now.

| Kardashev Scale | Barrow Scale | |
|---|---|---|
| KI – energy consumption at ~ 4 x $10^{19}$ erg s$^{-1}$ | BI – manipulates objects of its own scale | ~ 1 m |
| KII – energy consumption at ~ 4 x $10^{33}$ erg s$^{-1}$ | BII – manipulates genes | ~ $10^{-7}$ m |
| KIII – energy consumption at ~ 4 x $10^{44}$ erg s$^{-1}$ | BIII – manipulates molecules | ~ $10^{-9}$ m |
| | BIV – manipulates individual atoms | ~ $10^{-11}$ m |
| | BV – manipulates atomic nuclei | ~ $10^{-15}$ m |
| | BVI – manipulates elementary particles | ~ $10^{-18}$ m |
| | B$\Omega$ – manipulates space-time's structure | ~ $10^{-35}$ m |

**Table 1. Energetic and Inward civilization development**
Kardashev's (1964) types refer to energy consumption; Barrow's (1998, 133) types refer to a civilization's ability to manipulate smaller and smaller entities. In section 3, we combine those two scales.

## *1.1 Kardashev scale – the energetic increase*

Our civilization uses more and more energy. Energy is all-purpose, so we don't even need to understand the how or the why of this energy use to see that this trend is robust. Extrapolating this exponential increase of energy consumption, Kardashev (1964) showed that this would lead our civilization to type KII in year 5164 and to type KIII in 7764. Although Kardashev's original scale is an energetic one, it has often been interpreted as, and extrapolated to a spatial one. This is probably because the order of magnitude of the energy processed is as follows. Type KI harnesses the energy of a Earth-like planet; type KII harnesses the energy of a star and type KIII the energy of a galaxy. We are currently a ~KI civilization. Let us examine, as a typical example, our possible transition from type KI to type KII. What motivations could we have to harness the energy of the Sun? There are essentially two reasons. First, simply to meet our growing energy consumption needs; second, to avoid the predictable death of our Sun, associated with the destruction of life on Earth.

Let us first consider how to meet a civilization's growing energy needs. Einstein famously formulated the matter-energy equivalence formula $E=mc^2$. If we consider our solar system, where can we find most of its mass-energy? It is above all in the Sun, since 99.8% of our solar system's mass is in the Sun. That is, 99.8% of the energy in our solar system is to be found in the Sun. For any long-term use, the Sun is thus *the* obvious resource to harness energy from. Exploiting the energy of a star is an explorative engineering field known as *star lifting*, also called stellar mining, stellar engineering or asteroengineering (see e.g. Reeves 1985; Criswell 1985; Beech 2008).

The second incentive to engineer our Sun is to avoid its red giant phase which will begin in ~5 billion years. This enterprise is vital if we are concerned by saving life on Earth. Various processes have been proposed for this purpose, resulting in an elimination of this red giant phase. The topic is treated extensively by Martin Beech (2008). From a SETI perspective, this leads to concrete and observable predictions. Beech (2008, 190-191) indeed proposes 12 possible signs of stellar rejuvenation in progress.

## 1.2 Barrow scale – the inward manipulation

John Barrow (1998) classified technological civilizations by their ability to control smaller and smaller entities, as depicted in Table 1. This trend leads to major societal revolutions. Biotechnology, nanotechnology and information technologies are progressing at an accelerating pace and all stem from our abilities to control and manipulate small scales entities. This pivotal and overwhelming trend toward small spatial scales is largely overlooked in SETI, resulting in the Barrow scale being not well-known. Barrow estimates that we are currently a ~BIV civilization which has just entered nanotechnology.

Another argument for the importance of the Barrow scale is that, from the relative human point of view, there is more to explore in small scales than in large scales. As counter-intuitive as it is, space exploration offers more prospect in small scales than in large scales! This is illustrated in figure 1.

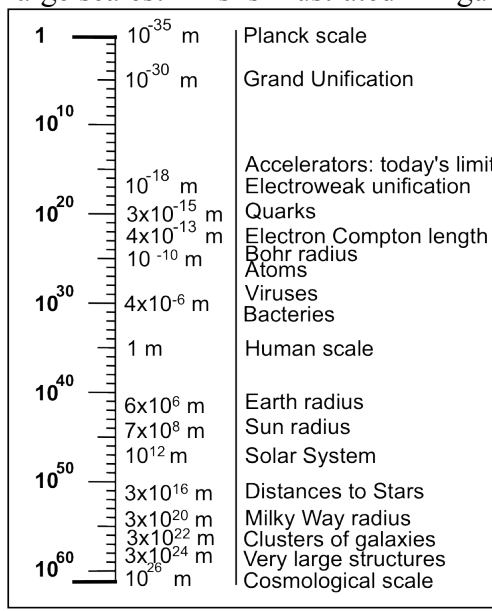

**Figure 1. Scales in the universe**
That humans are not in the center of the universe is also true in terms of scales. This implies that there is more to explore in small scales than in large scales. Richard Feynman (1960) popularized this insight when he said "there is plenty of room at the bottom". Figure adapted from (Auffray and Nottale 2008, 86).

In contrast with large cosmological scales, manufacturing, testing, exploring and exploiting small scale technologies is easier, cheaper and more controllable. It is also more efficient energetically. Since the development of such technologies is not hampered by the finiteness of the speed of light, its accelerating progress has no reason to slow down until we reach the Planck scale. Futurist and systems theorist John Smart (2009) characterized this trend as Space-Time-Energy-Matter (STEM) efficiency and density, or "STEM Compression". It can also simply be summarized with the motto of "doing more with less".

## 2 Black holes as attractors for intelligence

### 2.1 Unknown black hole technology

If we take seriously the Barrow scale, we do not even need to speculate on any particular black hole technology. We can simply assume that an intelligent civilization will develop to type BΩ, whatever its purpose is in using black holes. The reader averse to scientific and philosophical speculation might thus jump directly to section 3 for an application of the two-dimensional metric. However, black holes are fascinating attractors, not only because of their staggering gravitational field, but also because they are an intelligence's greatest potential. Let us see why with a short adventure on the speculative topic of black hole technology.

### 2.2 Energetic

Black holes are the densest objects in the universe. If we want to face the needs of consuming more energy, it might be beneficial to store or extract energy from black holes. Roger Penrose (1969, 270-272) imagined the following extraction mechanism. It consists of injecting matter into a black hole in a carefully chosen way, thereby extracting its rotational energy (see also Misner, Thorne, and Wheeler 1973, 908 for more details). Blandford and Znajek (1977) suggested a similar process with electrically charged and rotating black holes. Other proposals suggest collecting energy from gravitational waves of colliding black holes. Misner imagined this in 1968 as a personal communication to Penrose (1969). Frautschi (1982) also proposed to merge black holes as a way to produce a power source.

### 2.3 Societal

Which other societal functions could black holes fulfill? Louis Crane (2010, 370) has suggested that black holes are the perfect waste-disposal, although they should be manipulated with great care. He has also conducted an extensive study with Westmoreland on the possibility of black hole starships (Crane and Westmoreland 2009). Furthermore, general relativity leads to the fascinating topic of time travel via worm holes, theoretical cousins of black holes. Although no evidence of their existence is available, they could in theory provide shortcuts for traveling in space-time (for popular accounts see Thorne 1994; Randall 2005).

### 2.4 Scientific

Let us assume that terrestrial and ETIs are curious and continue to develop science. Black holes, especially their interiors, currently challenge our knowledge of the three fundamental physical theories: quantum mechanics, general relativity and thermodynamics. For scientific purposes, there might be an incentive to artificially produce black holes to better understand them. Indeed, since the 80's scientists have considered the possibility of making "universes in the lab" (see (see Ansoldi and Guendelman 2006 for a review). Although improbable sources of danger, some concerns have been raised regarding the accidental production of micro black holes in particle accelerators (Giddings and Thomas 2002). Still, we might want to produce them *intentionally* in the future.

A more concrete scientific application of black hole technology is to use them as telescopes or communication devices. How is it possible? An established consequence of general relativity theory is that light is bended by massive objects. This is known as *gravitational lensing*. For a few decades, researchers have proposed to use the Sun as a gravitational lens. At 22.45AU and 29.59AU we have a focus for gravitational waves and neutrinos. Starting from 550AU, electromagnetic waves converge. Those focus regions offer one of the greatest opportunity for astronomy and astrophysics, offering gains from 2 to 9 orders of magnitude compared to Earth-based telescopes. Over the years, Claudio Maccone (2009) has detailed with great technical precision such a scientific mission, called FOCAL. It is also worth noting that such gravitational lensing could also be used for communication. If we want to continue and improve our quest for understanding the cosmos, this mission is a great opportunity to complete our fuzzy astronomy with a focused one. In other words, the time may be ripe to put on our cosmic glasses.

But other ETIs may already have binoculars. Indeed, it is easy to extrapolate the maximal capacity of gravitational lensing using, instead of the Sun, a much more massive object, i.e. a neutron star or a black hole. This would probably consitute the most powerful possible telescope. This possibility was envisioned -yet not developed- by Von Eshleman in (1991). Since objects observed by gravitational lensing must be aligned, we can imagine an additional dilating and contracting focal sphere or artificial swarm around a black hole, thereby observing the universe in all directions and depths. Maybe such focal spheres are already in operation. The gains offered by such devices are currently unknown, but this is an exciting topic for an open minded researcher or a PhD student in general relativity.

## 2.5 Computational

What is the maximal information that can be processed by an advanced ETI? Visionary scientist Robert A. Freitas (1984) introduced the *sentience quotient*, which is a "scale of cosmic sentience universally applicable to any intelligent entity in the cosmos". At its limits, we have the maximal computational density of matter, what Seth Lloyd (2000) more recently called the "ultimate computer". What does such a computer look like? Lloyd argues that it is a black hole. Interestingly, if Moore's law is extrapolated, we attain such a maximal computational power by 2205 (Lloyd 2005, 162).

But black holes can be even more than ultimate computers. At the edge of theoretical computer science, some models of computation outperform Turing's original definition. Such devices are called *hypercomputers* (see e.g. Earman and Norton 1993). They are theoretically possible assuming particular space-time structures or with slowly rotating black holes (see e.g. Etesi and I Németi 2002; Andréka, I Németi, and P Németi 2009). If the construction of such hypercomputers is successful and indeed possible, this would bring qualitatively new ways to understand and model our universe. A breakthrough perhaps comparable to the invention of our ubiquitous computing machines.

## 2.6 Philosophical

Intelligence is the capacity to solve problems. It is by focusing on universal and long-term problems that we have the highest chances to understand the purpose of presumed ETIs. I see only two such serious problems. The first is the already mentioned red giant phase capable of wiping out life in a solar system like ours. This is a fundamental challenge any civilization born on the shore of a Sun-like star will have to face. A promising SETI strategy is thus to search for civilizations refusing this fate, by looking at artificially modified stars. According to Criswell (1985, 83) star lifting can considerably extend a civilization's time with matter to energy conversion, up to 2 millions times the present age of the universe, assuming the civilization stays at ~KI. Yet, even this runs out in the long term because the star will ultimately run out of usable energy.

What is the next level? Possibly migration, but that also cannot continue forever, because new star formation comes to an end in the very long term (Adams and Laughlin 1997). After realizing that the fate of stars is doomed, the longest term and truly universal problem is the continuation of the universe as a whole, to avoid its inevitable global entropy increase and death (see Ćirković 2003 for a review paper on physical eschatology).

The second challenge is thus, "How can we make life, intelligence and evolution survive indefinitely?" Answering this question is of course beyond the scope of this paper, but let us mention two proposals which include a role for black holes. Freeman Dyson proposed in his landmark (1979) paper that a civilization could hibernate and exploit the time dilation effects near black holes, to survive forever. However, this scenario doesn't work if the universe continues its accelerated expansion (Dyson 2004, xv). Yet, the core of the argument can be maintained if we replace digital computers by analog ones (see Dyson 2007).

Another speculative solution is to reproduce the universe (see e.g. Harrison 1995; Gardner 2003; Baláz 2005; Smart 2009; Gribbin 2009; Stewart 2010). This scenario combines the origin and future of the universe with a role for intelligent life. I have called it *Cosmological Artificial Selection* (see Vidal 2008, 2010; Vaas 2010, 2011 for critical commentaries; and Vidal 2011 for a response) as it is a philosophical extension of Smolin's (1992, 1997) theory of *Cosmological Natural Selection*. It is also worth noting that the future discipline of Artificial Cosmogenesis (Vidal 2008), analogous to Artificial Life but extended to the cosmos, would benefit the power of ultimate computers, to run simulations of whole universes.

Finally, if we assume that our universe is a black hole (e.g. Pathria 1972), the puzzling fine-tuning of universal constants could itself be interpreted as an intelligent signal from previous universe makers (Pagels 1989, 155-156; Gardner 2003). This is a radical proposal and the "Search for Extra Universal Intelligence" field has yet to emerge.

# 3 Black hole star lifting

## 3.1 The efficiency objection

In a SETI mindset, considering seriously that black holes are attractors for intelligence, we can now start to ask the following questions. What are the observable manifestations of a black hole when it's used as an energy source? as waste disposal? as a time-machine? as a starship engine? as an ultimate or hyper computer? as a universe production facility?

The exercise is highly speculative, and raises the *efficiency objection*. We saw that the Barrow scale trend makes civilizations develop with more and more efficiency. This would make small black holes more useful and thus hard or impossible to detect. It would be like trying to detect from Earth the existence of nanotechnology on the Moon. This is the essence of Smart's (2009) response to Fermi's paradox. We don't see other ETIs because they are confined inside black holes.

However, the two trends of more energy use and more energy efficiency need not be incompatible. Roughly speaking, our civilization has always been more efficient yet always using increasing amounts of energy. The key lies in the availability of energy. If it is poor, efficiency will strongly constrain civilization development. If energy is largely available, then efficiency matters less and civilizations can also grow on the Kardashev scale.

In his seminal paper, Dyson (1966, 643) assumed that ETIs would use technology we can understand. He qualified this assumption as "totally unrealistic". I do agree. However, there is a profound dilemma here. If we respect this rule, we restrict our search to civilizations roughly at our developmental level, not really higher. The search for ETIs *more advanced than us* is unlikely to succeed. But if we release this rule, -as I already did in section 2- this brings a paradox. Indeed, it will be hard, if not impossible, to argue that a phenomenon we don't understand is artificial, since its technology will, by definition, be alien to us. My answer for now is to give the two-dimensional metric a serious try.

To make SETI scientific, Dyson (1966) also writes that we should focus on the most conspicuous manifestations of intelligence and technology, so that we have something big to observe. I agree, and that time will follow Dyson's steps.

To summarize, on the Kardashev scale, we saw in section 1 that a type KII civilization would be able to use an amount of energy of the order of a star, with an endeavor called star lifting. Considering the magnitude of such an undertaking, it has good chances to be observable. On the Barrow scale, we have argued in section 2 that black hole technology attracts intelligence. We call a civilization with such technology, type B$\Omega$. It is the culmination of a civilization on that scale.

Now, can we derive a concrete SETI strategy, combining both the Kardashev and the Barrow scale? Could a civilization harness with great efficiency the energy of a star, to power its cutting edge black hole technology? Can we imagine to detect one day such a configuration?

## *3.2 Signs of KII-BΩ civilizations?*

We don't need to imagine or to wait because such configurations already exist! Indeed, 18 systems composed of a black hole accreting gas from a star have been found today (e.g. GRO J1655-40, GRS 1915+105, 1659-487, SS433, etc.). They are part of the family of binary systems, called X-Ray Binaries (XRB) because of their emissions in the X-Ray electromagnetic spectrum. Since a few decades, they are actively studied as natural -though sometimes intriguing- astrophysical systems. Importantly, researchers have concluded that a thin accretion disk around a rotating black hole is the *most efficient power source in the universe*, a process up to ~50 times more efficient than nuclear fusion occurring in stars (e.g. Thorne 1974; Narayan and Quataert 2005). If any civilization is to climb the Kardashev scale, it would certainly at some point want to master that energetic source. We call such an endeavor *black hole star lifting*.

Let us call such an hypothetical civilization KII-BΩ. It is of Type II on Kardashev's scale, because it is able to harness the energy of a star; and type Ω on Barrow's scale, since it manipulates space-time's structure with black hole technology. In fact, some XRB even display the main features of non-equilibrium systems. They have a strong energy flow from the star to the accreting black hole; at irregular intervals, plasma jets are ejected at relativistic velocities, which may be interpreted as entropy production; and, the black hole may be a structurally and informationally rich entity, if we assume that it could be a technology like an ultimate computer. We also can note that the black hole is *not* primarily used as an energy source, but as a technology which *needs* energy.

Accreting binaries are found in a great variety of configurations. Our two-dimensional metric allows to also speculate on the existence of other less advanced civilizations than KII-BΩ. A young type KII might be accreting the energy from a white dwarf star, which is nothing else than a burnt-out Sun-like star. A more mature KII may be using a neutron star accreting system. We can also hypothesize that some accreting neutron stars are in fact artificial black holes. Indeed, the main observational technique to decide whether the very dense accreting object is a neutron star or a black hole is to estimate its mass. If the object has a mass superior to Chandrasekhar's limit of 3 solar masses, it can only be a black hole. Otherwise, it is a neutron star. Therefore, the finding (by future methods) of a black hole less than 3 solar masses may corroborate its artificial origin.

What about more than KII? The Chandra X-Ray spatial telescope provided data showing that Low-Mass XRB (LMXB) are overabundant within 1 parsec of the galactic center (Muno et al. 2005). Could these be civilizations migrating toward the supermassive black hole? Although it sounds like a science-fiction novel, Vyacheslav I. Dokuchaev (2011) recently suggested that stable periodic orbits are theoretically possible inside supermassive black holes, and therefore, may be habitable.

Frank Tipler (1997) also envisioned with great details the possibility of a KΩ civilization mastering huge computational capacity. For various reasons, we will limit our discussion to the "modest" KII type.

## 3.3 Towards a new SETI agenda

Back to Earth, what can we do? We suggest to make a fresh start on XRB with a Dysonian and postbiological SETI perspective. Preparing this paper led me roughly to as many open questions as speculative ideas[1]. Let us mention just three important ones.

(1) "Can we find evidence of control in the XRB's energy flow?"
Energy flow regulation or control is a necessary condition for the growth, maintenance, evolution and reproduction of complex systems (see e.g. Aunger 2007; Chaisson 2011). Are some XRBs displaying such a feature?

(2) "How can we distinguish a natural process from an artificial and unknown one?"
We already mentioned the difficulty of such a task, but this is the price to pay to explore unknown technology. Accretion is an ubiquitous astrophysical process in galaxy and planetary formation, so XRBs may simply always be natural. Let me however introduce an analogy. Fission can be found in natural forms, as well as fusion, which is one of the core energetic processes in stellar evolution. Yet, humans try to copy them, and would greatly benefit to -always- control them. So it is not because a process is known to be natural that its actual use is not driven by an intelligence.
In fact, the situation may be even more subtle. The formation of XRBs might be natural, but controlled or taken over by ETIs, like a waterfall is a natural energy source humans can harness with dams. So, how can we develop criteria for natural versus artificial? Non-equilibrium thermodynamics, systems theory, artificial life, etc. because of their general concepts and applicability, will certainly provide key conceptual frameworks. Metrics like Freitas' *sentience quotient* or Chaisson's (2001, 2003) *energy rate density* are certainly very promising, and a KII-BΩ civilization should score high on them. Those metrics also indicate that the distinction natural versus artificial may be of a continuous nature.

Last but not least, (3) "What is the origin, evolution, fate and possible migration of XRB?" With this question I mean that a compelling argument for the existence of advanced ETIs will most likely come from an evolutionary and global cosmic understanding of natural and possibly artificial stellar evolution in our and other galaxies. Yes indeed, humanity's SETI is just getting started.

---

[1] Curious researchers are invited to the "Evo Devo Universe" community to brainstorm such questions at:
http://evodevouniverse.com/wiki/Research_on_SETI,_black_holes,_XRB_and_star_lifting_

# Conclusion

We have introduced in section 1 a two-dimensional metric for civilization development, Kardashev's scale assessing energy use and Barrow's scale assessing the ability to manipulate small scale technologies. At the Barrow scale limit, we find black hole technology. We substantiated this theoretical possibility by arguing that black holes are attractors for intelligence with energetic, scientific, societal, computational and philosophical arguments.

In section 3, we applied the two-dimensional metric, proposing that civilizations develop dramatically both their energy use and their small scale technologies. The limit for a KII civilization is to use the most efficient energy source known in the universe, accretion by a rotating black hole near a star. Combining those two limits, this leads to a civilization able to extract with great efficiency the energy of a star, while using black hole technology. Such systems have already been observed, and we hypothesize that they could be KII-BΩ civilizations. We thus invited SETI researchers to make a fresh start on X-Ray binaries, to assess the validity of this hypothesis.

The answer to "are we alone?" will certainly not come in a clear cut yes-no alternative. Proving that we are not alone will more plausibly progressively consolidate like the neo-darwinian theory of evolution, which stems from a convergence of observations of fossils, geological records, genetics, and careful study of millions of different living species. Our understanding of life in the universe is changing rapidly. Hundreds of exoplanets have been found, among them Earth-like planets susceptible to harbor life; if we are lucky, we may find meteorites containing alien bacteria, or even receive ETI signals; we have also good chances to observe conspicuous star lifting operations, and we need to assess the plausibility of some XRB being KII-BΩ civilizations.

Finally, let us remember the double-sided aspect of SETI. Assuming the worst-case SETI scenario, namely that we are truly alone in the universe, then all our speculations and insights may still prove very useful for our future (Dyson 1966). Whether we are alone or not, I hope this line of thinking will inspire humans and their descendants to continue both their energetic and inward development.

# Acknowledgments

I thank Martin Dominik and John Zarnecki for inviting me to present these ideas at the magnificent Kavli Royal Society International Centre. I was influenced by the work of John Smart, to take seriously the possibility of black hole technology; as much as by Freeman Dyson's seminal work on the possibility of stellar engineering. I thank David Brin, Georgi Georgiev, Francis Heylighen, Andreas Müller for helpful discussions.

# Appendix

Figure 2. maps the problem described in introduction, while figure 3. maps the core argument presented in the paper. Please read in a top-down direction. More details on argumentation mapping can be found in the Annex of (Vidal 2008).

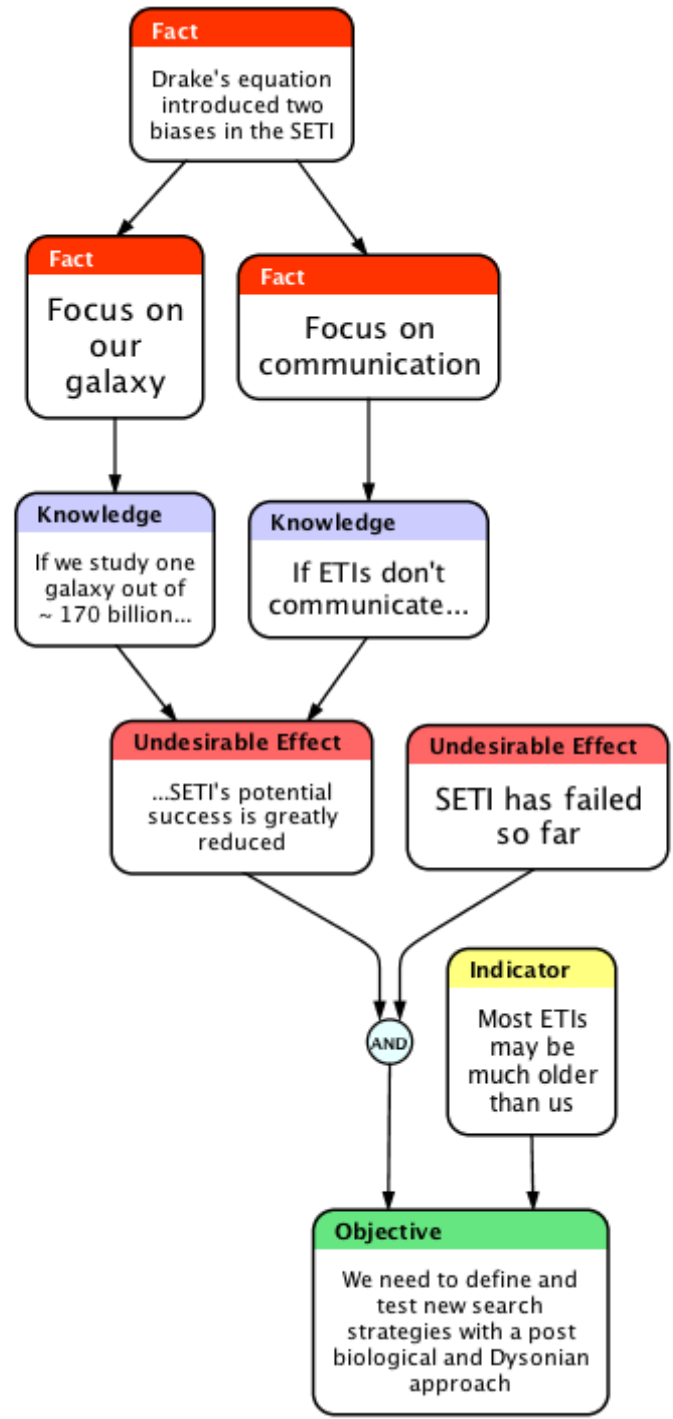

**Figure 2.** The current SETI situation described in the introduction

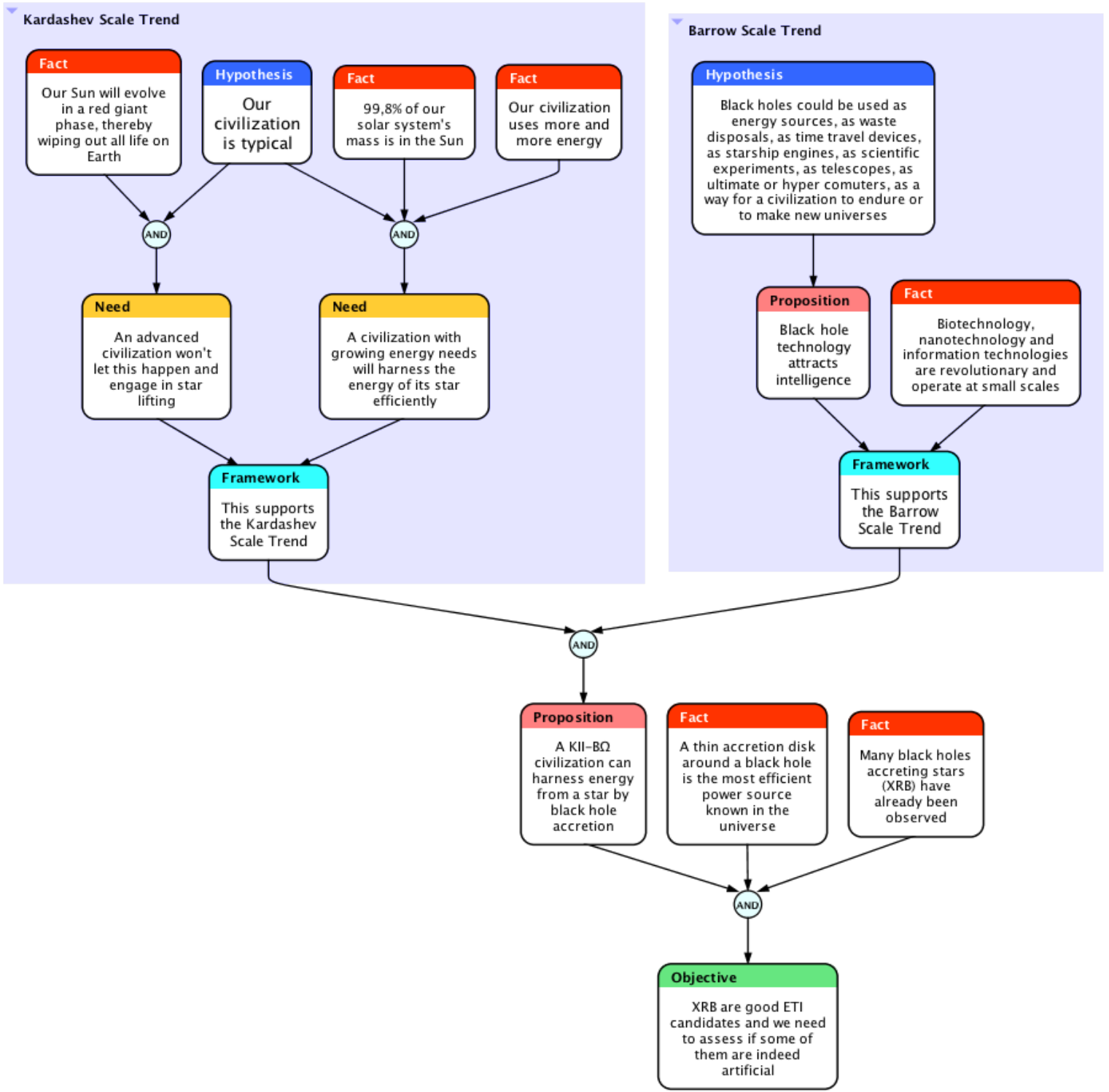

**Figure 3.** The main argument presented in the paper

# Bibliography

Adams, F. C., and G. Laughlin. 1997. "A Dying Universe: The Long-Term Fate and Evolution of Astrophysical Objects." *Reviews of Modern Physics* 69 (2): 337-372. http://arxiv.org/abs/astro-ph/9701131 .

Andréka, Hajnal, István Németi, and Péter Németi. 2009. "General relativistic hypercomputing and foundation of mathematics." *Natural Computing* 8 (3): 499-516. doi:10.1007/s11047-009-9114-3.

Ansoldi, Stefano, and Eduardo I Guendelman. 2006. "Child Universes in the Laboratory." *gr-qc/0611034* (November 5). http://arxiv.org/abs/gr-qc/0611034 .

Auffray, C., and L. Nottale. 2008. "Scale Relativity Theory and Integrative Systems Biology: 1 Founding Principles and Scale Laws." *Progress in biophysics and molecular biology* 97 (1): 79–114. http://fre2571.vjf.cnrs.fr/documentation/AUFFRAY-NOTTALE-PBMB.pdf .

Aunger, Robert. 2007. "A rigorous periodization of 'big' history." *Technological Forecasting and Social Change* 74 (8) (October): 1164-1178. doi:10.1016/j.techfore.2007.01.007.

Baláz, BA. 2005. "The Cosmological Replication Cycle, the Extraterrestrial Paradigm and the Final Anthropic Principle." *Diotima* (33): 44-53. http://astro.elte.hu/~bab/seti/IACP12z.htm .

Barrow, J. D. 1998. *Impossibility: The limits of science and the science of limits* . Oxford University Press, USA.

Beech, Martin. 2008. *Rejuvenating the sun and avoiding other global catastrophes* . Springer.

Blandford, R. D., and R. L. Znajek. 1977. "Electromagnetic extraction of energy from Kerr black holes." *Monthly Notices of the Royal Astronomical Society* 179: 433-456. http://adsabs.harvard.edu/abs/1977MNRAS.179..433B .

Chaisson, E. J. 2001. *Cosmic Evolution: The Rise of Complexity in Nature* . Harvard University Press.

———. 2003. "A Unifying Concept for Astrobiology." *International Journal of Astrobiology* 2 (02): 91-101. http://www.tufts.edu/as/wright_center/eric/reprints/unifying_concept_astrobio.pdf .

———. 2011. "Energy rate density as a complexity metric and evolutionary driver." *Complexity* 16 (3): 27-40. doi:10.1002/cplx.20323. http://www.tufts.edu/as/wright_center/eric/reprints/EnergyRateDensity_I_FINAL_2011.pdf .

Ćirković, Milan M, and Robert J. Bradbury. 2006. "Galactic gradients, postbiological evolution and the apparent failure of SETI." *New Astronomy* 11 (8) (July): 628-639. doi:10.1016/j.newast.2006.04.003. http://mcirkovic.aob.rs/paper_v4.pdf .

Ćirković, Milan M. 2003. "Resource Letter: PEs-1: Physical eschatology." *American Journal of Physics* 71: 122. http://www.aob.bg.ac.yu/~mcirkovic/Cirkovic03_RESOURCE_LETTER.pdf .

———. 2006. Macro-Engineering in the Galactic Context. In *Macro-Engineering*, ed. Viorel Badescu, Richard B. Cathcart, and Roelof D. Schuiling, 54:281-300. Dordrecht: Springer Netherlands. http://arxiv.org/abs/astro-ph/0606102 .

Crane, Louis. 2010. "Possible Implications of the Quantum Theory of Gravity: An Introduction to the Meduso-Anthropic Principle." *Foundations of Science* 15 (4): 369-373. doi:10.1007/s10699-010-9182-y. http://arxiv.org/abs/0912.5508 .

Crane, Louis, and Shawn Westmoreland. 2009. "Are Black Hole Starships Possible." *0908.1803* (August 12). http://arxiv.org/abs/0908.1803 .

Criswell, David R. 1985. Solar System Industrialization: implications for interstellar migrations. In *Interstellar Migration and the Human Experience* , ed. Ben R. Finney and Eric M. Jones, 50-87. University of California Press.

Dick, S. J. 2003. "Cultural evolution, the postbiological universe and SETI." *International Journal of Astrobiology* 2 (01): 65-74.

Dokuchaev, Vyacheslav I. 2011. "Is there life inside black holes?" *1103.6140* (March 31). http://arxiv.org/abs/1103.6140 .

Drake, F. D. 1961. "Project Ozma." *Physics Today* 14: 40.

Dyson, F. J. 1960. "Search for Artificial Stellar Sources of Infrared Radiation." *Science* 131 (3414): 1667 -1668. doi:10.1126/science.131.3414.1667.

———. 1966. The Search for Extraterrestrial Technology. In *Perspectives in Modern Physics*, ed. R.E. Marshak, 641–655. New York: John Wiley & Sons.

———. 1979. "Time Without End: Physics and Biology in an Open Universe." *Review of Modern Physics* 51: 447-460. http://www.think-aboutit.com/Misc/time_without_end.htm.

———. 2004. *Infinite in All Directions: Gifford Lectures Given at Aberdeen, Scotland April--November 1985*. Reprint. Harper Perennial.

———. 2007. *A Many-Colored Glass: Reflections on the Place of Life in the Universe*. Univ of Virginia Pr.

Earman, John, and John D. Norton. 1993. "Forever Is a Day: Supertasks in Pitowsky and Malament-Hogarth Spacetimes." *Philosophy of Science* 60 (1): 22-42.

Etesi, Gábor, and István Németi. 2002. "Non-Turing Computations Via Malament–Hogarth Space-Times." *International Journal of Theoretical Physics* 41 (2): 341-370-370. doi:10.1023/A:1014019225365. http://arxiv.org/abs/gr-qc/0104023.

Feynman, R. P. 1960. "There's plenty of room at the bottom." *Engineering and Science* 23 (5): 22–36.

Frautschi, S. 1982. "Entropy in an expanding universe." *Science* 217 (4560): 593–599.

Freitas, R. A. 1984. "Xenopsychology." *Analog Science Fiction/Science Fact* 104: 41-53. http://www.rfreitas.com/Astro/Xenopsychology.htm.

Gardner, J. N. 2003. *Biocosm. The New Scientific Theory of Evolution: Intelligent Life is the Architect of the Universe.* Inner Ocean Publishing.

Giddings, Steven B., and Scott Thomas. 2002. "High energy colliders as black hole factories: The end of short distance physics." *Physical Review D* 65 (5): 056010. doi:10.1103/PhysRevD.65.056010.

Gott III, J. R, M. Juric, D. Schlegel, F. Hoyle, M. Vogeley, M. Tegmark, N. Bahcall, and J. Brinkmann. 2005. "A Map of the Universe." *The Astrophysical Journal* 624: 463. http://arxiv.org/abs/astro-ph/0310571.

Gribbin, John. 2009. *In Search of the Multiverse*. Allen Lane.

Harrison, E. R. 1995. "The Natural Selection of Universes Containing Intelligent Life." *Quarterly Journal of the Royal Astronomical Society* 36 (3): 193-203. http://adsabs.harvard.edu/full/1996QJRAS..37..369B.

Kardashev, N. S. 1964. "Transmission of information by extraterrestrial civilizations." *Soviet Astronomy* 8 (2): 217–220. http://adsabs.harvard.edu/abs/1964SvA.....8..217K.

Lloyd, S. 2000. "Ultimate Physical Limits to Computation." *Nature* 406: 1047-1054. http://www.hep.princeton.edu/~mcdonald/examples/QM/lloyd_nature_406_1047_00.pdf.

———. 2005. *Programming the Universe: A Quantum Computer Scientist Takes on the Cosmos*. Vintage Books.

Maccone, Claudio. 2009. *Deep Space Flight and Communications: Exploiting the Sun as a Gravitational Lens*. Springer.

Misner, C. W, K. S Thorne, and J. A Wheeler. 1973. *Gravitation*. WH Freeman & co.

Muno, M. P., E. Pfahl, F. K. Baganoff, W. N. Brandt, A. Ghez, J. Lu, and M. R. Morris. 2005. "An Overabundance of Transient X-Ray Binaries within 1 Parsec of the Galactic Center." *The Astrophysical Journal* 622 (2): L113-L116. doi:10.1086/429721. http://arxiv.org/abs/astro-ph/0412492.

Narayan, Ramesh, and Eliot Quataert. 2005. "Black Hole Accretion." *Science* 307 (5706) (January 7): 77 -80. doi:10.1126/science.1105746. http://astron.berkeley.edu/%7Eeliot/science.pdf.

Pagels, Heinz R. 1989. *The Dreams of Reason*. New York: Bantam.

Pathria, R. K. 1972. "The Universe as a Black Hole." *Nature* 240 (5379): 298-299. doi:10.1038/240298a0.

Penrose, R. 1969. "Gravitational Collapse: The Role of General Relativity." *Riv. Nuovo Cim* 1: 252-276.

Randall, Lisa. 2005. *Warped passages unraveling the mysteries of the Universe's hidden dimensions*. New York: Ecco.

Reeves, H. 1985. *Atoms of Silence: An Exploration of Cosmic Evolution*. Cambridge, Massachusetts: MIT Press.

Shostak, Seth. 2010. "What ET will look like and why should we care." *Acta Astronautica* 67 (9-10): 1025-1029. doi:10.1016/j.actaastro.2010.06.028.

Smart, J. 2009. Evo Devo Universe? A Framework for Speculations on Cosmic Culture. In *Cosmos and Culture: Cultural Evolution in a Cosmic Context*, ed. S. J. Dick and Mark L. Lupisella, 201-295. Washington D.C.: Government Printing Office, NASA SP-2009-4802. http://accelerating.org/downloads/SmartEvoDevoUniv2008.pdf .

Smolin, L. 1992. "Did the Universe evolve?" *Classical and Quantum Gravity* 9 (1): 173-191.

———. 1997. *The Life of the Cosmos*. Oxford University Press, USA.

Stewart, John E. 2010. "The Meaning of Life in a Developing Universe." *Foundations of Science*. doi:10.1007/s10699-010-9184-9. http://cogprints.org/6655/ .

Thorne, K. S. 1974. "Disk-Accretion onto a Black Hole. II. Evolution of the Hole." *The Astrophysical Journal* 191: 507-520. http://adsabs.harvard.edu/abs/1974ApJ...191..507T .

———. 1994. *Black Holes and Time Warps: Einstein's Outrageous Legacy*. WW Norton & Company.

Tipler, Frank J. 1997. *The Physics of Immortality: Modern Cosmology, God and the Resurrection of the Dead*. Anchor.

Vaas, R. 2010. "Life, the Universe, and almost Everything: Signs of Cosmic Design?" *0910.5579*. http://arxiv.org/abs/0910.5579 .

———. 2011. "Cosmological Artificial Selection: Creation out of something?" *Foundations of Science*. doi:10.1007/s10699-010-9218-3. http://arxiv.org/abs/0912.5508 .

Vidal, C. 2008. The Future of Scientific Simulations: from Artificial Life to Artificial Cosmogenesis. In *Death And Anti-Death*, ed. Charles Tandy, 6: Thirty Years After Kurt Gödel (1906-1978).:285-318. Ria University Press. http://arxiv.org/abs/0803.1087 .

———. 2010. "Computational and Biological Analogies for Understanding Fine-Tuned Parameters in Physics." *Foundations of Science* 15 (4): 375-393. doi:10.1007/s10699-010-9183-x. http://arxiv.org/abs/1002.3905 .

———. 2011. "Fine-tuning, Quantum Mechanics and Cosmological Artificial Selection." *Foundations of Science*. doi:10.1007/s10699-010-9219-2. http://arxiv.org/abs/0912.5508 .

Von Eshleman, R. 1991. Gravitational, plasma, and black-hole lenses for interstellar communications. In *Bioastronomy The Search for Extraterrestial Life — The Exploration Broadens*, 390:299. Lecture Notes in Physics. Springer Berlin / Heidelberg. http://dx.doi.org/10.1007/3-540-54752-5_236 .